

\def\title{
\centerline{NEW APPLICATIONS OF}\vskip 1mm
\centerline{THE BIEDENHARN-TEMPLE OPERATOR
\foot{Dedicated to Prof. L. C. Biedenharn, on his 70th 
birthday.
{\it Festschrift in honor of L. C. Biedenharn}, ed. B. Gruber. Plenum :
New York (1994).}
}}

\def\runningtitle{The Biedenharn-Temple operator}

\def\author{\vfill
\centerline{P. A. HORV\'ATHY}
\vskip 3mm
\centerline{D\'epartement de Math\'ematiques,
Universit\'e,}
\centerline{Parc de Grandmont,
 F-37200 TOURS (France)}
\centerline{e-mail: horvathy@univ-tours.fr}
\vfill
}

\def\runningauthor{
Horv\'athy}


\vsize = 9.2truein
\hsize = 6.4truein 
\baselineskip = 12 pt 

\headline ={
\ifnum\pageno=1\hfill
\else\ifodd\pageno\hfil\tenit\runningtitle\hfil\tenrm\folio
\else\tenrm\folio\hfil\tenit\runningauthor\hfil
\fi\fi
} 

\nopagenumbers
\footline = {\hfil} 


\def\parag{\hfil\break} 
\def\IR{{\bf R}} 
\def\ccr{\cr\noalign{\medskip}}
\def\smallcirc{{\raise 0.5pt\hbox{$\scriptstyle\circ$}}}
\def\and{\qquad \hbox{and}\qquad}

\def\2{{1\over 2}}
\def\D{D\llap{\big/}}

\font\tenb=cmmib10 
\newfam\bsfam

\textfont\bsfam=\tenb

\mathchardef\alphab="080B
\mathchardef\betab="080C
\mathchardef\omegab="0821
\mathchardef\lambdab="0815
\mathchardef\xib="0818
\mathchardef\etab="0811
\mathchardef\pib="0819
\mathchardef\sigmab="081B


\newcount\ch 
\newcount\eq 
\newcount\foo 
\newcount\ref 

\def\chapter#1{
\parag\eq = 1\advance\ch by 1{\bf\the\ch.\enskip#1}
}

\def\equation{
\leqno(\the\ch.\the\eq)\global\advance\eq by 1
}

\def\foot#1{
\footnote{($^{\the\foo}$)}{#1}\advance\foo by 1
} 

\def\reference{
\parag [\number\ref]\ \advance\ref by 1
}

\ch = 0 
\foo = 1 
\ref = 1 

\title
\vskip 3mm
\author
\vskip 2mm
\parag{\bf Abstract.}

{\it The Biedenharn approach to the Dirac-Coulomb problem is applied to a system
considered by D'Hoker and Vinet, which consists of a spin $\2$ particle in the
combined field of a Dirac monopole plus a $\lambda^2/r^2$
potential. The explicit solution is obtained by diagonalizing the 
'Biedenharn-Temple operator', $\Gamma$}.

\chapter{Introduction.}

In a paper anticipating
supersymmetric quantum mechanics, [1], Biedenharn proposed a
new approach to the Dirac-Coulomb problem. He starts with the square of
the Dirac equation. This equation can be written in a  non-relativistic
Coulomb form (but with a fractional angular momentum), by using 
the \lq Biedenharn-Temple' [1, 2] operator
$\Gamma$.  
The solutions of the first-order equation can be recovered 
from those of the second-order equation
by projection.

Recently [3], we have applied Biedenharn's method to the
\lq dyon' system of D'Hoker and Vinet [4]. Here we illustrate the 
Biedenharn method on yet another
example (also considered by D'Hoker and Vinet [5]), namely that of
 a spin $\2$
particle described by the four-component Hamiltonian
$$
H =\pmatrix{H_1&\cr &H_0\cr} = 
\2\bigg\{\pib^2 -q{\sigmab.\hat{\bf r}\over r^2} +{\lambda^2\over r^2}
-\lambda\gamma^5{\sigmab.\hat{\bf r}\over r^2}\bigg\}
\equation
$$
where $\lambda$ is a real constant.

\chapter{The Biedenharn-Temple operator for a monopole system.}

The Hamiltonian (1.1) can be viewed as associated to a static gauge field 
on $\IR^4$,
${\bf A}=q{\bf A}_D$,
$A_4=\lambda/r$,
where ${\bf A}_D$ denotes the vector potential of a Dirac
monopole of unit strength. The square of the
associated Dirac operator,
$$
\D:=\pmatrix{&Q^{\dagger}\cr Q&\cr} = 
\pmatrix{&\sigmab.\pib - i{\lambda\over r}\ccr
\sigmab.\pib + i{\lambda\over r} &\cr},
\equation
$$
is seen to be (1.1). Since the Dirac operator is
chiral-supersymmetric in any even dimension, the partner hamiltonians have the
same spectra. 

The conserved total angular
momentum is 
$
{\bf J}={\bf L}+\sigmab/2,
$ 
where
$
{\bf L} = {\bf \ell} - q{\hat{\bf r}}$, 
${\bf \ell} = {\bf r}\times {\bf\pib}.
$ The eigenvalues of ${\bf J}^2$ are 
$j(j+1)$, $j=q-1/2,\ q+1/2,\ldots$.
Set   
$$
w =\sigmab.{\hat {\bf r}},\qquad z ={\bf\sigmab.\ell} + 1 
\and {\cal K} = -\pmatrix{&iz\cr iz&\cr}.
\equation
$$
Note that $w^2 = 1$, and that $z$
anticommutes with $w$ and ${\bf\sigmab.\pib}$,  
$$
\lbrace z, w\rbrace = 0
\and
\lbrace z, 
{\bf\sigmab.\pib} \rbrace = 0.
\equation
$$
${\cal K}$ commutes with
the Dirac operator $\D$. Using $({\bf \sigmab.L})^2 = {\bf L}^2 +
i{\bf\sigmab}({\bf L}\times{\bf L})={\bf L}^2 -{\bf\sigmab.L}$, 
one proves that 
$
{\cal K}^2\,=\,z^2\,=\,{\bf J}^2 + 1/4-q^2.
$
Thus, $z$ (and $\cal K$) have {\it irrational} eigenvalues,  
$$
\kappa =\pm\sqrt{(j+1/2)^2-q^2}\ . 
\equation
$$ 
Since $j\geq q -
1/2$, ${\cal K}$ is hermitian, but for $j = q -1/2$ its eigenvalue $\kappa$
vanishes and ${\cal K}$ is not invertible.
 
Using $\sigmab.\pib = -iw(\partial_r+1/r - z/r)$,
the supercharges $Q$ and
$Q^{\dagger}$ can be written as 
$$\left\{\matrix{ 
&Q = &-iw\Big(\partial_r + {\displaystyle 1\over\displaystyle r} 
- {\displaystyle y\over\displaystyle r}\Big) &=
&-i\Big(\partial_r + {\displaystyle 1\over\displaystyle r} 
+ {\displaystyle x\over\displaystyle r}\Big) w, 
\ccr
&Q^{\dagger} = &-iw\Big(\partial_r +{\displaystyle 1\over\displaystyle r} 
-{\displaystyle x\over\displaystyle r}\Big)
&= &-i\Big(\partial_r + {\displaystyle 1\over\displaystyle r} 
+ {\displaystyle y\over\displaystyle r}\Big)w,}\right. 
\equation
$$
where we introduced the notations
$$
x=z-\lambda w
\and
y=z+\lambda w.
\equation
$$
($xw = - wy$).
Let us define the {\it Biedenharn-Temple operator} as
$$
\Gamma = - ({\bf\sigmab.\ell}+1+\gamma^5\lambda
w) 
\qquad\hbox{i.e.}\qquad
 - (z+\gamma^5\lambda w)\equiv -\pmatrix{y&\ccr&x}.
\equation
$$
Since
$
\Gamma^2= z^2 +\lambda^2={\bf J}^2+1/4 +\lambda^2-q^2,
$
the eigenvalues of $\Gamma$ are  
$$
\gamma =\pm\sqrt{(j+1/2)^2+\lambda^2-q^2},
\qquad
\hbox{sign}\ \gamma =\hbox{sign}\ \kappa.
\equation
$$
In terms of $\Gamma$, $\D^2$ becomes 
$$
\D^2 = 
\pmatrix{Q^{\dagger} Q \cr &Q Q^{\dagger}\cr}
= - (\partial _r + {1\over r})^2 + 
{\Gamma(\Gamma +1)\over r^2}.
\equation
$$
Although $\Gamma$ does not commute with $\D$, it is conserved for the
quadratic dynamics.

\chapter{The explicit solution.}

The operator $\Gamma$ can be diagonalized generalizing
the procedure in Ref. 6. Let us first assume that $j\geq
j+1/2$, and let $L_{\pm} := j\pm1/2$ be the orbital angular momentum quantum
number. Consider first the two-component spinors   
$$
\varphi_{\pm}^{\mu} =\sqrt{{L_{\pm} + 1/2\pm\mu\over 2L_{\pm} + 1}}\, 
Y^{\mu-1/2}_{L_{\pm}}\pmatrix{1\ccr 0\cr} 
\pm 
\sqrt{{L_{\pm} + 1/2\mp\mu\over 2L_{\pm} + 1}}\, 
Y^{\mu +1/2}_{L_{\pm}}
\pmatrix{0\ccr 1\cr},
\equation 
$$
where the $Y$'s are the monopole harmonics and the
sign $\pm$ refers to the sign of $\gamma$. The $\varphi$'s 
satisfy
$
{\bf J}^2= j(j+1), 
\;
J_3 =\mu,\ (\mu = -j,\cdots,j),
\;
{\bf L}^2
 = L_{\pm}(L_{\pm}+1).
$ 
The $2$-spinors
$$
\left\{\matrix{
&\chi_+&={1\over 2j+1}\,\Big(&\varphi_+&+ &{q\over {j+1/2 +\mid\kappa\mid}}
\ \varphi_- &\Big)
\ccr
&\chi_-&={1\over 2j+1}\,\Big(&- 
{q\over {j+1/2 +\mid\kappa\mid}} 
\ \varphi_+&+ &\varphi_- &\Big)\cr}\right.  
\equation
$$
satisfy the important relations
$$
z\ \chi_{\pm} = \pm\vert\kappa\vert\ \chi_{\pm}
\and
w\ \chi_{\pm} = \chi_{\mp},
\equation
$$
as well as  
${\bf J}^2=j(j+1)$, 
$J_3=\mu,\ (\mu = -j,\cdots, j)$. 
Hence
$$\left\{
\matrix{
\phi_+=(\vert\kappa\vert+j+{1\over 2})\ \chi_+-\lambda\ \chi_-,
&\quad\phi_-=\,\lambda\ \chi_+ +(\vert\kappa\vert+j+{1\over 2})\chi_-
\ccr
\Phi_+=(\vert\kappa\vert+j+{1\over 2})\ \chi_++\lambda\ \chi_-,
&\quad\Phi_-=-\lambda\ \chi_+ +(\vert\kappa\vert+j+{1\over 2})\chi_-
\cr}\right.
\equation
$$
diagonalize $x$ and $y$,
$$
x\phi_{\pm}^{\mu} =\mp\mid\gamma\mid\phi^{\mu}_{\pm}
\and
y\Phi_{\pm}^{\mu} =\mp\mid\gamma\mid\Phi^{\mu}_{\pm}.
\equation
$$

The operator $w =\sigmab.{\hat{\bf r}}$ interchanges
the $x$ and $y$ eigenspinors, 
$$
w\,\phi_{\pm}^{\mu}=\Phi^{\mu}_{\mp}\ .
\equation
$$

For $j=q-1/2$, no $\varphi_-$ is available and $\chi_-$ is hence missing.
$\chi_+$ is proportional to the lowest $\varphi_+$ in (3.1). Thus, there are no 
$\phi_-$-states in the $\gamma^5 = -1$ sector and no $\Phi_+$ states
in the $\gamma_5 = 1$ sector. However, in each $\gamma^5$ sector, (3.2) 
yields $(2q)$ states, namely
$$
{\phi_+^\mu}^0 = {\Phi_-^\mu}^0 \propto\sqrt{{q+1/2\pm\mu\over 2q + 1}}\, 
Y^{\mu-1/2}_q\pmatrix{1\ccr 0\cr} 
\pm 
\sqrt{{q+1/2\mp\mu\over 2q + 1}}\, 
Y^{\mu +1/2}_q\pmatrix{0\ccr 1\cr} 
\equation 
$$
They are $(+1)$-eigenstates of $w$.

The eigenfunctions of $\D^2$ are then found as 
$$\matrix{
&\left\{
\eqalign{&
\Psi_{\pm\mid\gamma\mid} = u_{\pm} 
\pmatrix{\Phi_{\pm}\cr 0\cr} 
\ccr 
&\psi_{\pm\mid\gamma\mid} = u_{\pm} 
\pmatrix{0\cr\phi_{\pm}}
\cr} 
\right.
\qquad\hbox{for}\quad
\left\{\matrix{\gamma^5 = 1\cr\ccr\cr
\gamma^5 = - 1\cr}\right.
&\hbox{for}\;j\geq q+1/2\ccr
&\left\{
\eqalign{&
\Psi_-^0 = u_-^0 
\pmatrix{\Phi_-\cr 0\cr} 
\ccr 
&\psi_+^0 = u_+^0
\pmatrix{0\cr\phi_+\cr}
\cr} 
\right.
\qquad\hbox{for}\quad
\left\{\matrix{\gamma^5 = 1\cr\ccr\cr
\gamma^5 = - 1\cr}\right.
&\hbox{for}\;j= q-1/2
 \cr}
\equation
$$

Thus, the radial functions $u_{\pm}(r)$ solve the equations 
$$
\Big[-(\partial_r +{1\over r})^2 +{\gamma
(\gamma +1)\over r^2}-2E\Big]\ u_{\pm} = 0.  
\equation
$$
This is the wave equation for a free particle except for 
the fractional \lq angular momentum' $\gamma$. Its solutions is 
hence given by the Bessel functions,
$$
u_{\pm}(r)\,\propto\,r^{-1/2}\
J_{\vert\gamma\vert\mp\2}(\sqrt{2E}\ r).
\equation
$$

For $\lambda = 0$ we recover the formulae in [6]. The well-known 
self-adjointness problem in the $j=q-1/2$ sector shows up in that
 the eigenvalue $\gamma$ in Eq. (2.8) vanishes in this case. 
(Self-adjointness of $\D^2$ requires in fact $\vert\lambda\vert\geq 3/2$ [5]).

Another particular value is $\lambda =\pm q$, when the Biedenharn-Temple 
operator has half-integer eigenvalues,
$$
\gamma =\pm (j+\2).
\equation
$$
In this case, $\gamma(\gamma+1)$ is the same for
 $-\vert\gamma\vert$ as for $\vert\gamma\vert -1$,
leading to identical solutions for (3.9). 
Thus, the corresponding energy levels
are two-fold degenerate. (This only happens for  
$\vert\gamma\vert\geq\vert\gamma\vert_{min}+1$ i.e. for
$j\geq q+1/2$).

This can also be understood by noting that, for
$\lambda =\pm q$, the spin dependence drops out in 
one of the $\gamma^5$-sectors. For $\lambda = q$, e.g., the Hamiltonian 
(1.1) reduces to
$$
H =\pmatrix{H_1&\ccr &H_0\cr} = 
\2\,\pmatrix{
\pib^2 -2q{\displaystyle\sigmab.\hat{\bf r}\over\displaystyle
 r^2} +{\displaystyle q^2\over\displaystyle r^2}&\ccr
&\pib^2 +{\displaystyle q^2\over\displaystyle r^2}\cr
},
\equation
$$
i.e.  $H_0$ describes a spin $0$ particle, while
$H_1$ corresponds to a particle with anomalous gyromagnetic ratio $4$. 
The system admits hence an extra $o(3)$ symmetry, 
generated by the spin vectors
$$
\matrix{
{\bf S}_0= \2\sigmab
&\qquad
{\bf S}_1=U^{\dagger}{\bf S}_0U
\ccr
\hbox{for} \;H_0
&\qquad\hbox{for} \;H_1\cr},
\equation
$$
where
$
U = Q/\sqrt{H_1}$ 
and  
$U^{-1} = U^{\dagger} = 
1/\sqrt{H_1}\ Q^{\dagger}
$
are the unitary transformations which intertwine the non-zero-energy parts of 
the chiral sectors. 

Each of the partner Hamiltonians $H_1$ and $H_0$ in (1.1) have a 
 non-relativistic conformal $o(2,1)$ symmetry [7] which
combines, together with $\D$ and $-i\gamma^5\D$, 
into an $osp(2,1)$ superalgebra [5]. Since the Hamiltonian (2.9) is essentially
the same as for a free particle, the question arises whether the $osp(2,1)$
likely extends into the super-Schr\"odinger algebra [8]. 

For further results the Reader is referred to [9].


\vskip3mm
\centerline{\bf References}

\reference
L. C. Biedenharn, Phys. Rev. {\bf 126}, 845 (1962); 
L. C. Biedenharn and N. V. V. Swamy, Phys. Rev. {\bf 5B}, 1353 (1964);
M. Berrondo and H. V. McIntosh, 
Journ. Math. Phys. {\bf 11}, 125 (1970).

\reference
G. Temple, {\it The General Principles of Quantum Mechanics}, New York: Methuen (1948);
P. C. Martin and R. J. Glauber, 
Phys. Rev. {\bf 109}, 1307 (1958).

\reference
F. Bloore and P. A. Horv\'athy,
Journ. Math. Phys. {\bf 33}, 1869 (1992).

\reference
E. D'Hoker and L. Vinet, 
Phys. Rev. Lett. {\bf 55}, 1043 (1986);
E. D'Hoker, V. A.
Kostelecky and L. Vinet, in {\it Dynamical Groups and Spectrum Generating
Algebras}, p. 339-367, World Scientific, Singapore (1988).

\reference
E. D'Hoker and L. Vinet, 
Comm. Math. Phys. {\bf 97}, 391-427 (1985).

\reference
Y. Kazama, C. N. Yang and A. S. Goldhaber, Phys. Rev. {\bf D15}, 2287 (1977);
Y. Kazama and C. N. Yang, Phys. Rev. {\bf D15}, 2300 (1977).

\reference
R. Jackiw, Ann. Phys. (N.Y.) {\bf 129}, 183 (1980).

\reference
J. P. Gauntlett, J. Gomis and P. K. Townsend, Phys. Lett. {\bf 248B}, 288 
(1990); C. Duval and P. A. Horv\'athy, {\it Journ. Math. Phys.} {\bf 35}, 2516 (1994).

\reference
P. A. Horv\'athy, A. J. Macfarlane and J.-W. van Holten,
 Phys. Lett. {\bf B 486}, 346-352 (2000).

\bye